\documentclass[aps,prx,notitlepage,superscriptaddress,showpacs,twocolumn ]{revtex4-2}
\usepackage{graphicx,subfigure,epsfig}
\usepackage{array,multirow}
\usepackage{dcolumn}
\usepackage{amssymb,amsmath,amsfonts,mathrsfs}
\usepackage{array}
\usepackage{times,setspace}
\usepackage{latexsym}
\usepackage{float,flafter,bm,bbm}
\usepackage{epstopdf,color,multirow}
\usepackage[colorlinks,linkcolor=blue,anchorcolor=blue,urlcolor=blue,citecolor=blue]{hyperref}
\usepackage{footnote}
\usepackage[OT1]{fontenc}

\usepackage[normalem]{ulem}

\hypersetup{
   colorlinks=true,
   linkcolor=blue,
   filecolor=magenta,
   urlcolor=blue,
}

\begin{document}

\title{{{ Proposal for asymmetric photoemission and tunneling spectroscopies in quantum simulators of the triangular-lattice Fermi-Hubbard model}}}

\date{\today}
\author{Shuai A. Chen}
\affiliation{Department of Physics, Hong Kong University of Science and Technology, Clear Water Bay, Hong Kong, China}
\author{Qianqian Chen}
\affiliation{Kavli Institute for Theoretical Sciences, University of Chinese Academy
of Sciences, Beijing 100190, China}
\author{Zheng Zhu}
\email{zhuzheng@ucas.ac.cn}
\affiliation{Kavli Institute for Theoretical Sciences, University of Chinese Academy of Sciences, Beijing 100190, China}
\affiliation{CAS Center for Excellence in Topological Quantum Computation, University of Chinese Academy of Sciences, Beijing, 100190, China}
\date{\today}

\begin{abstract}
Recent realization of well-controlled quantum simulators of the triangular-lattice Fermi-Hubbard model, including the triangular optical lattices loaded with ultracold Fermions and the heterostructures of the transition-metal dichalcogenides, as well as the more advanced techniques to probe them, pave the way for studying frustrated Fermi-Hubbard physics. Here, we theoretically predict asymmetric photoemission and tunneling spectroscopies for a lightly hole-doped and electron-doped  triangular Mott antiferromagnet, and reveal two distinct types of magnetic polarons: a \emph{lightly} renormalized quasiparticle with the same momentum as the spin background and a \emph{heavily} renormalized quasiparticle with a shifted momentum and a nearly flat band, using both analytical and unbiased numerical methods. We propose these theoretical findings to be verified in frustrated optical lattices and  Moir\'e superlattices by probing various observables including  the spectral function, the density of states, the energy dispersion and the quasiparticle weight. Moreover, we reveal the asymmetric response of the spin background against charge doping, demonstrating that the interplay between the local spin and charge degrees of freedom plays a vital role in doped triangular Mott antiferromagnets.
\end{abstract}

\maketitle


\section{Introduction}
The Fermi-Hubbard model is widely believed to be a prototypical model to capture the essential physics of many realistic strongly correlated systems, notably the doped Mott insulators~\cite{WenDopingRMP}. Important insights into the doped Mott insulators could be gained by investigating the motion of the doped single charge and its interplay with the spin background. Due to its potential relevance to the high $T_c$ cuprates~\cite{Anderson1987,Keimer2015}, such an issue has been extensively studied on the square lattice~\cite{Dagotto1994} and the interplay between spin and charge has been proved to play an essential role~\cite{Brinkman1970Single,Shraiman1988Mobile,
Kane1989Motion,Horsch1991spinpolaron,Zhu2013Strong,2015PhRvB92w5156Z,2016NuPhB.903...51Z,PhysRevB.98.035129,CS2019Singlehole,PhysRevB.98.245138,2021arXiv210614898Z,Fabian2020parton, Fabian2021rotational}.
Nevertheless, those analogous problems on the triangular lattice, which are equally important and likely to exhibit distinct physics due to the geometric frustrations and the absence of particle-hole symmetry, still need plenty of endeavors. Recently, quantum simulating the Fermi-Hubbard model has been realized in both cold-atom optical lattices~\cite{Gross2021,Bohrdt2021, Bakr2009quantumgas, Parsons2016siteresolved, Sherson2010singleatomresolved, Cheuk2016Observation,YangJin2021,Schafer2020, Boll2016,Brown2017, Brown2019, GuardadoSanchez2020,Koepsell2020,Hartke2020, Bloch2012,Wurz2018,Chiu2019,Koepsell2019,2020NatPh1626B,Bohrdt2018,Lewenstein2012} and condensed-matter Moir\'{e} superlattices~\cite{Tang2020Simulation,Kennes2021platform}, and the theoretical predictions can thereby be verified.

\begin{figure}[!h]
\centering
\includegraphics[scale=0.5]{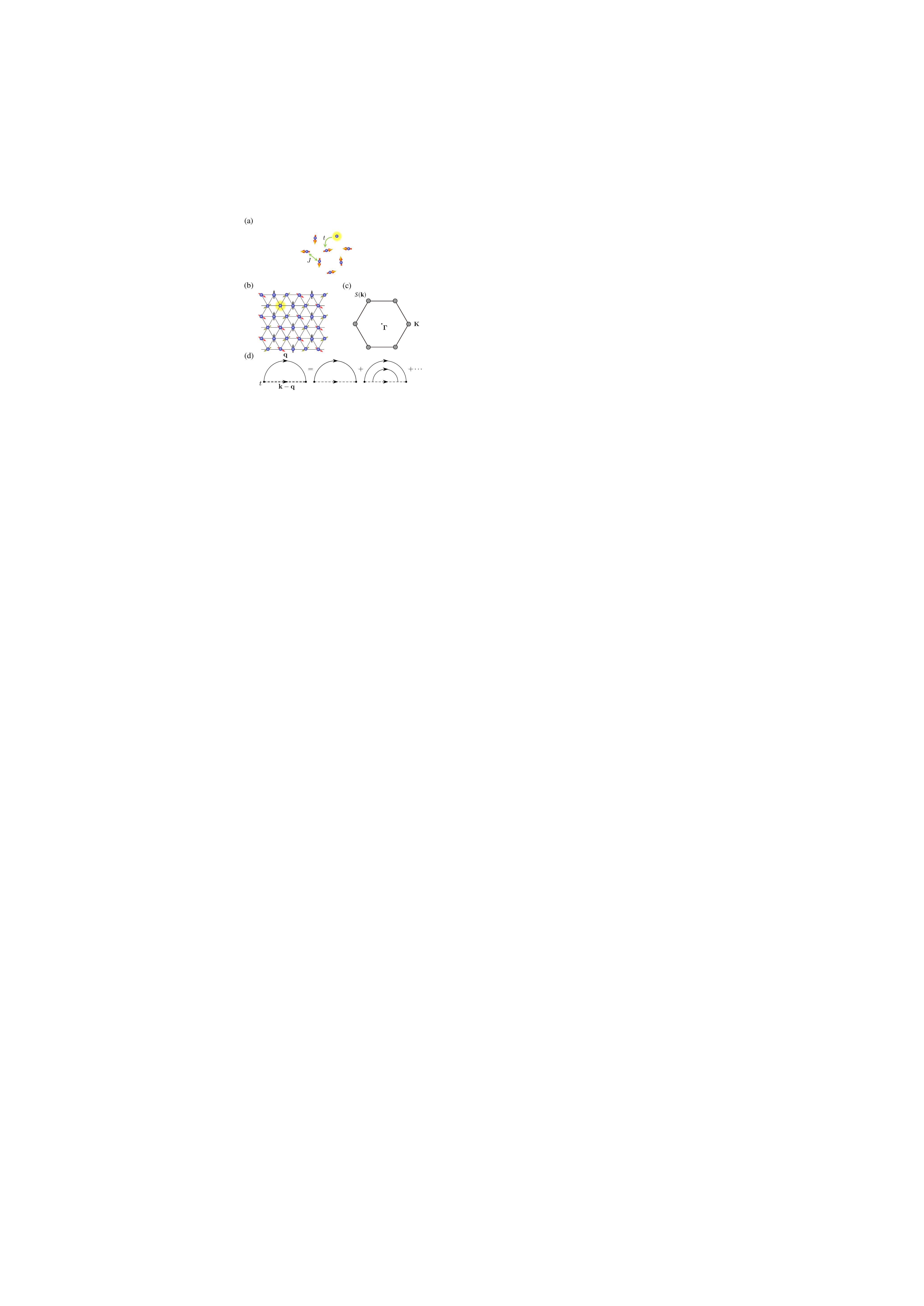}
\renewcommand{\figurename}{Fig.}
\caption{(Color online.) (a) Fermions trapped in a triangular optical lattice. A $120^\circ$ antiferromagnetic order emerges with a characteristic energy scale $J$, and it further dresses a bare hole with a cloud of magnons.
(b) Schematic diagram of triangular Mott antiferromagnet with a single doped charge.
(c) Schematic diagram of the static spin structure factor $S(\mathbf{k})$ for the $120^\circ$ N\'eel order at half filling. The maxima of $S(\mathbf{k})$ are denoted by solid gray circles.
(d) The rainbow diagram for SCBA calculations. Holon and magnon are presented by solid and dashed lines, respectively. The double dashed line represents the exact holon propagator. The vertex marked by the black dot is of order $t$.
}
\label{Fig:fig1}
\end{figure}

Compared with the complicated condensed-matter materials~\cite{Tang2020Simulation,Kennes2021platform},
the optical lattice, which is a more controlled and pristine platform, has been remarkably advanced
\cite{Lewenstein2012,2020NatPh1626B,Bohrdt2018,Wurz2018,Bloch2012, Gross2021,Bohrdt2021,Bakr2009quantumgas, Parsons2016siteresolved, Sherson2010singleatomresolved, Cheuk2016Observation,YangJin2021,Schafer2020,Boll2016,Brown2017, Brown2019,GuardadoSanchez2020,Koepsell2020,Hartke2020,Chiu2019,Koepsell2019}, where the momentum-resolved spin structure factor \cite{Wurz2018}, the spectral function \cite{2020NatPh1626B,Bohrdt2018} and the real-space motion of doped holes \cite{Chiu2019,Koepsell2019} can be probed using high-resolution techniques.
In particular, more recently, the triangular optical lattice loaded with ultracold Fermions has been implemented experimentally \cite{YangJin2021}, as illustrated in Fig.~\ref{Fig:fig1}(a), where the coupling strength $U$ and the hopping amplitude $t$ could be accurately tuned through Feshbach resonance \cite{ChinCheng2010} and the strength of the optical lattice \cite{Lewenstein2012}, respectively.
It is possible to precisely track the motion of a single charge and obtain its interplay with fluctuating spin backgrounds \cite{Chiu2019,Koepsell2019} due to a revolutionary real-space detection using quantum microscopy~\cite{Gross2021} that images the dynamics of all ultracold atoms simultaneously and the angle-resolved photoemission spectroscopy (ARPES) that measures spectral function \cite{2020NatPh1626B,Bohrdt2018}.

In a different context, the transition-metal dichalcogenide (TMD) and its heterostructures, such as WSe$_2$/WS$_2$ hetero-bilayers~\cite{Tang2020Simulation},
provide a distinct platform to simulate the triangular-lattice Hubbard model~\cite{Tang2020Simulation,Kennes2021platform} with widely tunable parameters like the charge carrier density~\cite{Pan2018Quantum-Confined, 2019arXiv191012147W,Regan2020Mott,An2020Interaction}, and advanced measurement techniques, such as the spectral functions and density of states by nano-ARPES~\cite{Lisi2021} and scanning tunneling microscope (STM).

 Nevertheless, it is highly controversial for previous studies on the fate of injected charges in a triangular Mott
 antiferromagnet ~\cite{PhysRevB53402,Trumper2004Quasiparticle,Vojta1999spinpolaron,PhysRevB72224409,2022arXiv220203458K}
 even for those with the same method like self-consistent Born approximation (SCBA)~\cite{PhysRevB53402,Trumper2004Quasiparticle} (details see Appendix).
Therefore, resolving  this critical problem is a much-needed task both theoretically and experimentally.
Motivated by the above aspects, in this paper, we propose the photoemission and tunneling spectroscopies of 
lightly hole-doped and electron-doped
triangular Mott antiferromagnet.
By establishing the magnetic polaron theory analytically with self-consistent Born approximation (SCBA), we theoretically predict the asymmetric photoemission and tunneling spectroscopies with respect to the particle-hole transformation, and identify two distinct types of magnetic polarons: a lightly renormalized quasiparticle with the same momentum as the spin background and a heavily renormalized quasiparticle with a shifted momentum and nearly flat band,
which implies even richer physics as compared to the square lattice.
The density matrix renormalization group (DMRG) simulation, which resolves the sharp ambiguity from small size effect \cite{PhysRevB53402,Trumper2004Quasiparticle} in the exact diagonalization (ED) method, numerically backs up the analytical conclusions. We further show the asymmetric responses of the spin background against charge doping concentration and demonstrate the validity of our theory at light doping. We remark that the signature of the asymmetry proposed here, including the spectral function, the density of states, and the static spin structure factor, can be directly probed in recently realized frustrated optical lattices and TMD hetero-bilayers.

\begin{figure}[t]
\centering \includegraphics[scale=0.4]{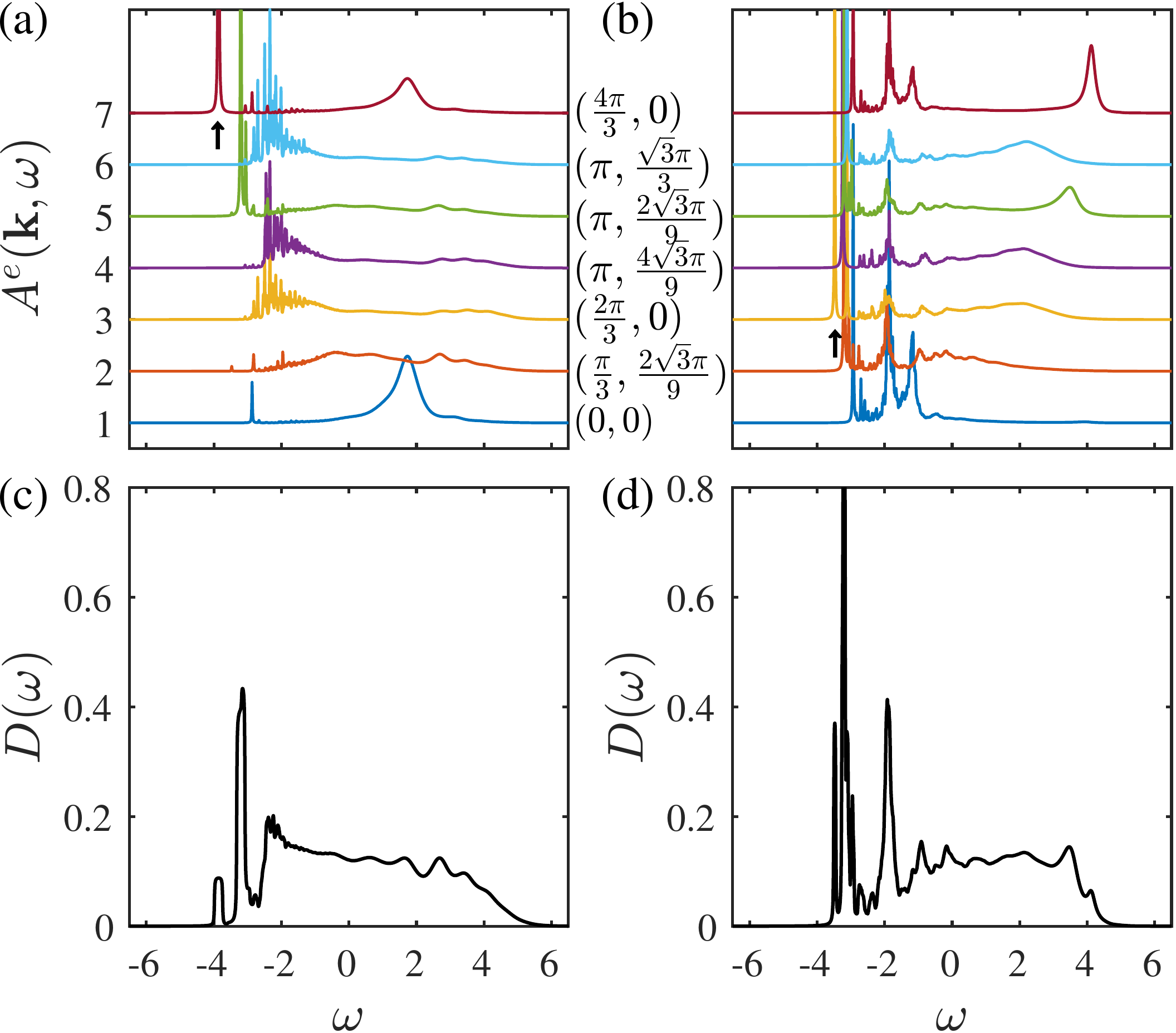}
\renewcommand{\figurename}{Fig.}
\caption{(Color online.)
The spectral function $A^{e}(\mathbf{k},\omega)$ [(a) and (b)]
    and density of states $D(\omega)$ [(c) and (d)] obtained from SCBA at $N=6\times 6$ lattice with  periodic boundary conditions.
     A well-defined quasiparticle peak (marked by black arrows) lies at the bottom of spectrum with $\mathbf{K}=(\frac{4\pi}{3},0)$ for $t/J=5$ in (a) and $(\frac{2\pi}{3},0)$ for $t/J=-5$ in (b), respectively.
    The density of states $D(\omega)$ features a vanishing gap in a continuum limit for both sides,
    and for $t/J=-5$ in (d), it shows the divergent 
    behavior. Here we set $|t|=1$ as the unity of energy, and $\Delta\omega=0.01$.
}
\label{Fig:AkwAw}
\end{figure}

\section{Model Hamiltonian}
The motion of the doped charge in a triangular lattice can be described by the Fermion-Hubbard model
\begin{equation}
H = -t \sum_{\langle ij\rangle \sigma} (c_{i\sigma}^\dagger c_{j\sigma} +h.c.)+ U \sum_i
n_{i\uparrow}n_{i\downarrow}~,
\label{model}
\end{equation}
where ${c_{i\sigma}^{\dagger}}$($c_{i\sigma}$) and $n_i$ denote a fermion creation (annihilation)  and particle number operators at site $i$, respectively.  The summation runs over all
the nearest-neighbor links $\langle ij\rangle$.
In this work, we focus on the strong coupling regime, where
the Hamiltonian at half-filling reduces to a pure Heisenberg spin model with the superexchange coupling $J=4t^2/U$ and $120^{\mathrm{\circ}}$ N\'eel order in the ground state
\cite{Sachdev1992Kagome,Fa2006spinliquid,Chernyshev2009spinwaves,Song2019unifying}. Upon doping a charge, i.e., $\sum_{i}n_{i}=N-1$ ($N$ the number of the lattice sites),
the hopping process is triggered with amplitude $t$. Then the low-energy effective Hamiltonian reads
\begin{equation}
 H=-t\mathcal{P}\sum_{\left\langle ij\right\rangle \sigma}(c_{i\sigma}^{\dagger}c_{j\sigma}+h.c.)\mathcal{P}+J\sum_{\left\langle ij\right\rangle } \mathbf S_i\cdot \mathbf S_j~,
\end{equation}
where $\mathcal P$ projects to the single-occupancy subspace
i.e. $n_i\leq 1$ and $\mathbf S_i$ is a spin operator $\mathbf S_i=\frac{1}{2}c_i^\dagger(\sigma_x,\sigma_y,\sigma_z)c_i$ with $\sigma_{x,y,z}$ being Pauli matrices and $c_i=[c_{i\uparrow},c_{i\downarrow}]^T$. Unlike the bipartite
square-lattice case~\cite{Brinkman1970Single,Shraiman1988Mobile,
Kane1989Motion,Horsch1991spinpolaron,Zhu2013Strong,Zhu2014Nature, 2015PhRvB92w5156Z,CS2019Singlehole,PhysRevB102104512,PhysRevB.98.245138,2021arXiv210614898Z, Fabian2020parton, Fabian2021rotational},
the particle-hole symmetry is absent here with the physics depending on the sign of hopping amplitude $t$~\cite{Wang2004Dopedtj}.
We therefore perform a comparative study for $t>0$ and $t<0$, which can be connected by a full particle-hole transformation for both spins (see Appendix). We are devoted to light doping exemplified by a single charge doping with the methods of the SCBA \cite{Kane1989Motion,Horsch1991spinpolaron,PhysRevB53402,Trumper2004Quasiparticle} and DMRG \cite{Whitedmrg,Ostlund1995Thermodynamic}.

\section{Magnetic polaron theory}
We start from the half-filled spin background with an in-plane order, here the $120^{\mathrm{\circ}}$ N\'eel order is characterized by momentum $\mathbf{Q=K}$ for the triangular Mott antiferromagnets~\cite{Sachdev1992Kagome,Fa2006spinliquid, Chernyshev2009spinwaves,Song2019unifying}. We consider the order in XZ-plane and an in-plane rotation of each spin at $\mathbf{r}_{i}$ by an angle $\mathbf{Q}\cdot \mathbf{r}_{i}$.
Implementing the Holstein-Primakoff transformation with $S_{i}^{z}=S-a_i^\dagger a_i$, $S_{i}^{+}=a_{i}$, we introduce a boson $a$ to describe the low-energy magnon excitations and get the Hamiltonian $H_{a}=\sum_{\mathbf k}\omega_{\mathbf k}^{s}\beta_{\mathbf k}^{\dagger}\beta_{\mathbf k}^{}$, where $\omega_{\mathbf k}^{s}=\frac{\nu}{2}JS\sqrt{\left(1-\gamma_{\mathbf k}\right)\left(1+2\gamma_{\mathbf k}\right)}$ denotes the energy dispersion of magnons and the canonical modes $\beta_{\mathbf k}^{}=u_{\mathbf k}^{}a_{\mathbf k}^{}-\upsilon_{\mathbf k}^{}a_{-\mathbf k}^{\dagger}$. Here $\gamma_{\mathbf{k}}=\sum_{\delta}e^{i\mathbf{k}\cdot\delta}/{\nu}$ sums over all the $\nu=6$ nearest-neighbor sites on the triangular lattice and $u_{\mathbf k},\upsilon_\mathbf{k}$ are usual Bogoliubov factors in spin-wave theory. Then the rotated ground state can be constructed as $\vert\Psi_{0}\rangle=\exp(-\sum_{\mathbf{k}}\frac{v_{\mathbf{k}}} {u_{\mathbf{k}}}a_{\mathbf{k}}a_{-\mathbf{k}})\vert \mathrm{N\text{\'e}el}\rangle$.

Upon doping, the motion of the charge is dressed by magnons, forming a magnetic polaron.
The creation of a spinless charge is described by $h_{i}^{\dagger}$ in the fractionalization scheme: $c_{i\uparrow}^{}=h_{i}^{\dagger}$ and $c_{i\downarrow}^{}=h_{i}^{\dagger}S_{i}^{+}= h_{i}^{\dagger}a_{i}$.
Ignoring higher-order interactions, we derive the effective Hamiltonian with two terms: the kinetic energy term
\begin{equation}
H_{h0}=  -\sum_{\mathbf{k}}\omega^h_{0}(\mathbf{k})h_{\mathbf{k}}^{}h_{\mathbf{k}}^{\dagger}
\end{equation}
with dispersion relation
$\omega^h_{0}(\mathbf{k})=\frac{\nu t}{2}(\gamma_{\mathbf{k}+\frac{\mathbf Q}{2}}+\gamma_{\mathbf{k}-\frac{\mathbf Q}{2}})$ and
the holon-magnon coupling term
\begin{equation}
H_{hb}=-\frac{\nu t}{i\sqrt{N}}\sum_{\mathbf{k},\mathbf q }h_{\mathbf{k}}^{}h_{\mathbf{k}-\mathbf{q}}^{\dagger} \left(M_{\mathbf{k}} a_{\mathbf{q}}^{\dagger}-M_{\mathbf{k-q}}a_{-\mathbf{q}}\right)
\end{equation}
 with $M_{\mathbf{k}}=\gamma_{\mathbf{k}+\frac{\mathbf{Q}}{2}}-\gamma_{\mathbf{k}-\frac{\mathbf{Q}}{2}}$, which describes the motion of the charge in the process of absorbing or emitting magnons.
We remark that 
the kinetic energy term of the magnetic polarons is absent in the square-lattice case~\cite{Kane1989Motion,Horsch1991spinpolaron}. To obtain the single-particle Green's function $G^{h}(\mathbf{k},\omega)\equiv\langle\Psi_{0}\vert h_{\mathbf k}\frac{1}{\omega-H}h_{\mathbf k}^{\dagger}\vert\Psi_{0}\rangle$, we adopt the SCBA, i.e, considering the rainbow Feynman diagrams sketched in Fig.~\ref{Fig:fig1}(d), to get the self energy
$$
\Sigma^h(\mathbf k, \omega)\! =\! \sum_\mathbf{q}\frac{f(\mathbf k,\mathbf q)}{\omega -\omega_0^h(\mathbf k-\mathbf q)-\omega_\mathbf{q}^s-\Sigma^h(\mathbf k-\mathbf q,\omega-\omega_\mathbf{q}^s)},
$$
where a vertex coupling $f(\mathbf{k},\mathbf{\mathbf{q}})=(\nu t)^{2}\vert M_{\mathbf{k}}^{}\upsilon_\mathbf{q}-M_{\mathbf{k}-\mathbf{q}}u_{\mathbf q}\vert ^{2}/{N}$
originates from holon-magnon interaction.

\begin{figure}[t]
    \centering \includegraphics[scale=0.42]{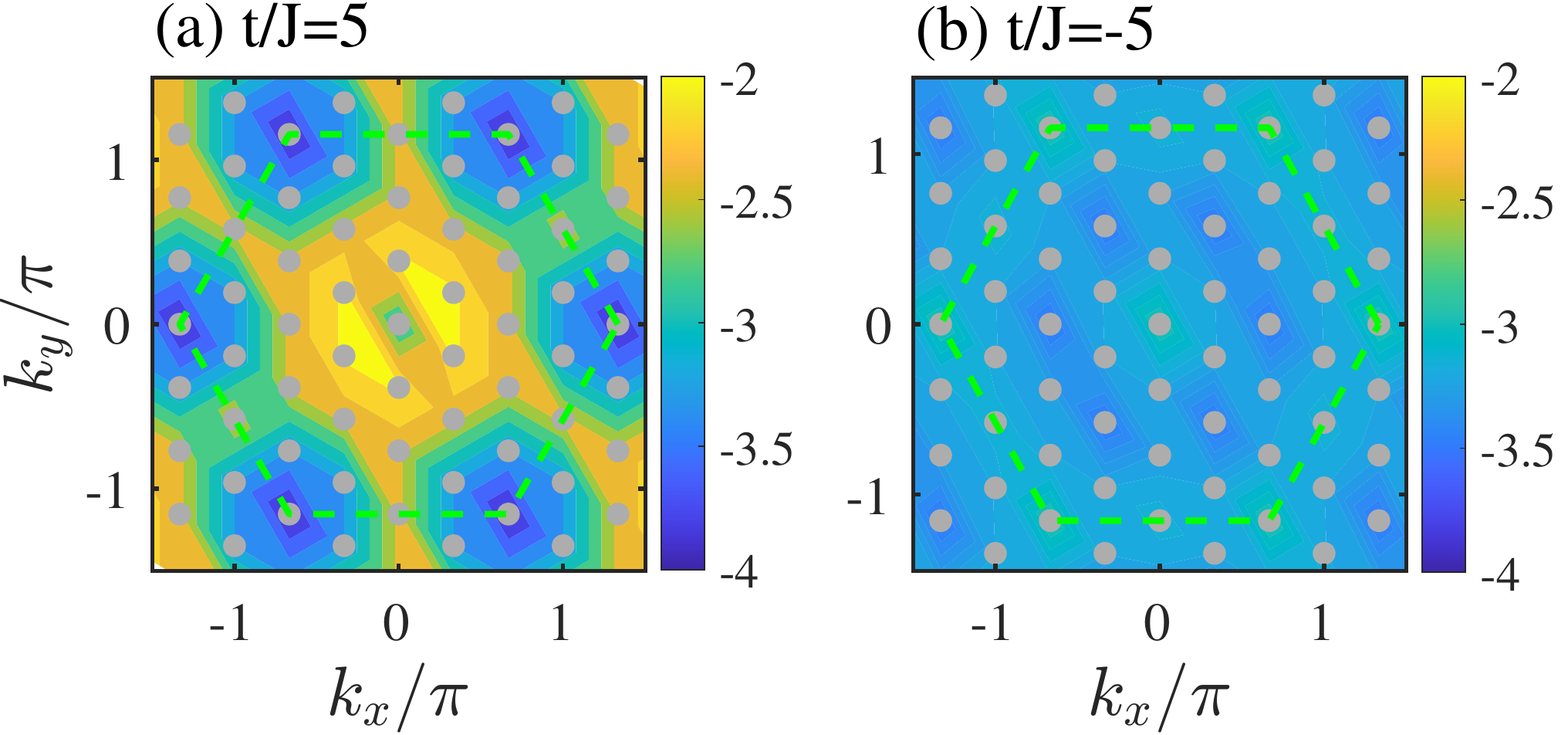}
    \renewcommand{\figurename}{Fig.}
    \caption{
        (Color online.) {Energy dispersion.} Energy dispersion for a single charge in (a) $t/J=5$ and (b) $t/J=-5$ from SCBA. The gray dots mark the accessible Bloch momenta on a $N=6\times 6$ lattice.
   }
    \label{Fig:dispersion}
\end{figure}

\section{Photoemission and Tunneling Spectroscopies}
To directly compare with the experiments, we compute the Green's function $G^{e}(\mathbf{k},\omega)$ with a relation to $G^{h}(\mathbf{k},\omega)$ via
\begin{equation}
G^{e}(\mathbf{k},\omega)=-\frac{1}{4} G^{h}(\mathbf{k}+\frac{\mathbf{Q}}{2},\omega) -\frac{1}{4}G^{h}(\mathbf{k}-\frac{\mathbf{Q}}{2},\omega)~.
\end{equation}
The momentum shift $\pm\mathbf{Q}/2$ is induced by the rotation on $\vert\Psi_{0}\rangle$, and for triangular Mott antiferromagnets, $\mathbf{Q=K}$.

We will show the spectral function $A^e(\mathbf k,\omega)$ [see Figs.~\ref{Fig:AkwAw}(a-b)] and the density of states $D(\omega)$[see Figs.~\ref{Fig:AkwAw}(c-d)], both of which are experimentally detectable through the ARPES \cite{2020NatPh1626B,Bohrdt2018} and scanning tunneling microscope (STM). The spectral functions $A^{e}(\mathbf{k},\omega)\equiv\frac{1}{\pi} \operatorname{Im}G^{e}(\mathbf{k},\omega+i\delta)$ are presented in Figs.~\ref{Fig:AkwAw}(a-b) for $t/J=\pm 5$ on a $N=6\times 6$ lattice with energy resolution $\Delta \omega$. We have confirmed their robustness on the larger size with the convergence guaranteed by energy resolution $\Delta \omega=0.01$ (see Appendix).
For $t/J=5$, as shown in Fig.~\ref{Fig:AkwAw}(a), we find that, in the spectral function, the sharp peak at momentum $\mathbf{K}=(\frac{4\pi}{3},0)$ with the lowest energy signals a well-defined quasiparticle, which separates from the peaks at other momenta with higher energy.
Interestingly, unlike the square-lattice case, where the $180^{\circ}$ N\'eel order and the doped hole locate at momenta $(\pi,\pi)$ and $(\pi/2,\pi/2)$, respectively, here the momentum of the doped charge is the same as that of the spin background with $120^{\circ}$ N\'eel order.
However, for $t/J=-5$, although we can still identify a well-defined quasiparticle peak at momentum $(\frac{2\pi}{3},0)$, there are numerous excitations with a fairly close energy scale,  as demonstrated in Fig.~\ref{Fig:AkwAw}(b), suggesting a heavily reduced bandwidth compared with the $t/J=5$ side. This observation demonstrates that the doped charge is highly renormalized with much larger effective mass and much smaller quasiparticle weight.
The spectral function for $t/J>0$ and $t/J<0$ manifests the asymmetric photoemission spectroscopies.
Notably, we can observe some Lorentz-like broadening peaks at much higher energies with momenta $(\frac{4\pi}{3},0)$ and $(0,0)$, which reflects the spin dynamics and confirms our theoretical setup.

Moreover, we compute the density of states $D(\omega)=\sum_{\mathbf{k}}A^{e}(\mathbf{k},\omega)$ that can be directly probed by STM. As shown in Fig.~\ref{Fig:AkwAw}(c),
$D(\omega)$  is also asymmetric with respect to $t/J>0$ and $t/J<0$.
We find $D(\omega)$ characterizes a well-defined quasiparticle when $t/J>0$, however, for $t/J<0$, the divergence of $D(\omega)$ near the ground state signals the Van Hove singularity behavior or the nearly flat band [see Fig.~\ref{Fig:AkwAw}(d)], suggesting the numerous excitations near the ground state, consistent with many low-energy peaks in a narrow energy window in the  spectral functions $A^{e}(\mathbf{k},\omega)$ [see Fig.~\ref{Fig:AkwAw}(b)].

\begin{figure}[t]
    \centering
    \includegraphics[width=0.5\textwidth]{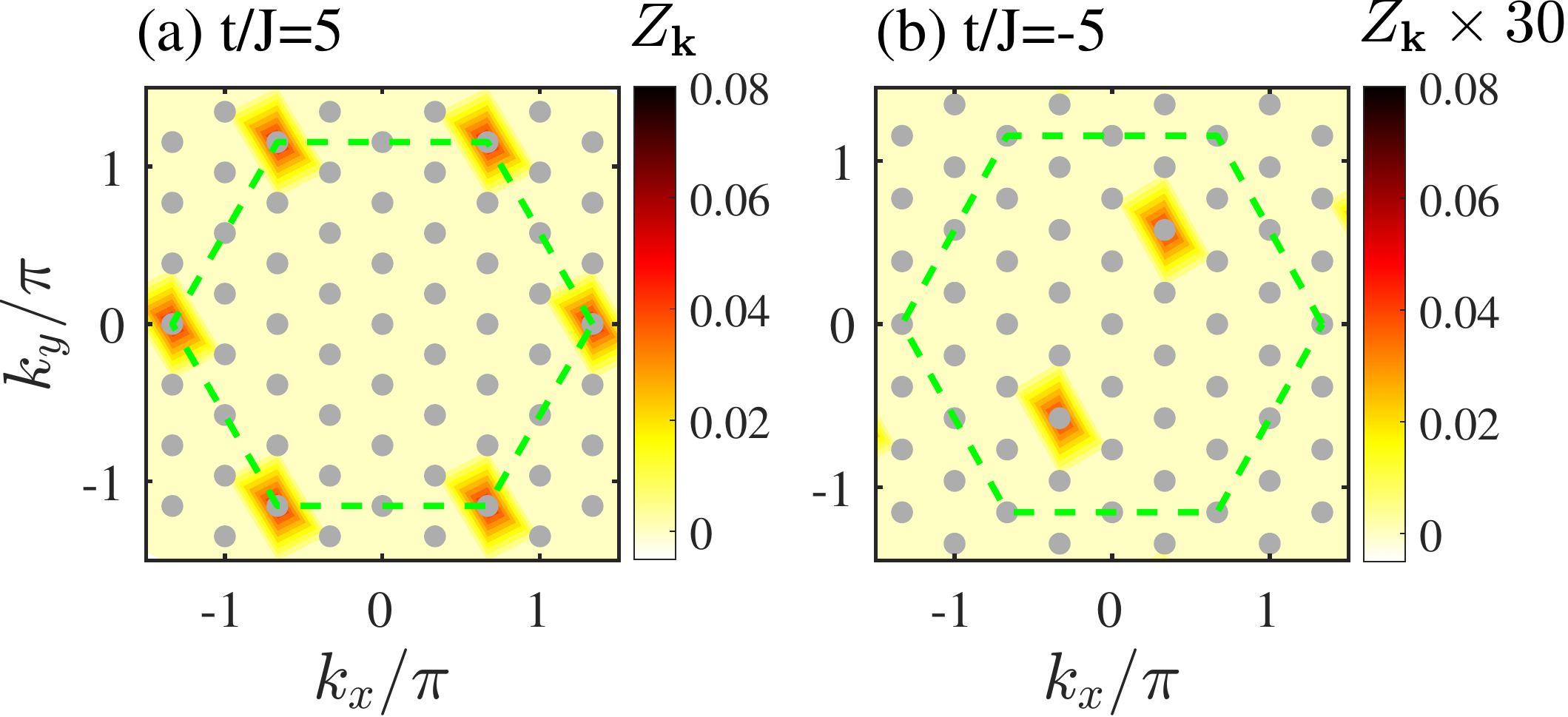}
    \renewcommand{\figurename}{Fig.}
    \caption{(Color online).
    The quasiparticle spectral weight $Z_\mathbf{k}$ calculated by DMRG for (a) $t/J=5$ and (b) $t/J=-5$ with a single charge doping.
    The gray dots represent the accessible momenta with $N=6\times6$ cylinder geometry.
    }
    \label{Fig:ZkDMRG}
\end{figure}

We remark that the asymmetric spectral functions $A^e(\mathbf k,\omega)$ and density of states $D(\omega)$ with respect to $t/J>0$ and $t/J<0$ indicate the distinct bandwidth for the doped charge, which can be inferred from the gaps between peaks in $A^e(\mathbf k,\omega)$ [see Figs.~\ref{Fig:AkwAw}(a) and (b)], or more clearly, from the energy dispersion $\omega^h(\mathbf k)$ shown in Fig.~\ref{Fig:dispersion}. Within the same color scale, we can find the nearly flat band with vanishingly small bandwidth when $t/J<0$, but a well-defined dispersive quasiparticle when $t/J>0$.
Further, consistency is shown at a larger lattice size (see Appendix).
The nearly flat band implies the doped charge is heavily renormalized, giving rise to much larger effective mass or much smaller quasiparticle weight. The minimum in the energy dispersion $\omega^h(\mathbf k)$ characterizes the ground-state momentum of the doped charge, which is also confirmed by the DMRG simulations below.

\section{Numerical simulation}
Below we employ DMRG to confirm the validity of SCBA by examining the quasiparticle weight. In addition to the single charge doping, we further examine the finite doping to understand the asymmetric behavior by probing the spin background against doping and further reveal that our proposal works at light doping.
Numerically it requires the integral multiple of $3$ for system length $L_x$ and width $L_y$ in order to accommodate the $120^{\mathrm{\circ}}$ N\'eel order, and DMRG computational cost increases exponentially with $L_y$, we therefore primarily focus on $L_y=6$ cylinders with the bond dimension up to D=20000. The charge doping can be accurately controlled by implementing U(1) symmetries.

We compute the quasiparticle spectral weight distribution $Z_\mathbf{k}$, which is defined by the overlap between the one-charge doped ground-state wave function $\Psi_{\text{1-charge}}$ and the wave function obtained by removing a charge from the half-filled ground state $\Psi_{\text{0-charge}}$, i.e.,
\begin{equation}
Z_\mathbf{k}\equiv |\langle \Psi_{\text{1-charge}}|c^{}_\mathbf{k}|\Psi_{\text{0-charge}} \rangle |^2~.
\end{equation}
Figures~\ref{Fig:ZkDMRG}(a) and (b) show the contour plot of $Z_\mathbf{k}$ for $t/J=5$ and $t/J=-5$ on $L_y=6$ cylinders, respectively.
The finite $Z_\mathbf{k}$ for both $t/J>0$ and $t/J<0$ suggest well-defined quasiparticles, consistent with the nature of magnetic polarons.
The location of the peaks in $Z_\mathbf{k}$ characterizes the ground-state momentum of the doped charge. For $t/J=5$, the significant peaks locate at momenta $\mathbf{K}$ [see Fig.~\ref{Fig:ZkDMRG}(a)], while for $t/J=-5$, we find that the peaks locate at momentum $\mathbf{K}/2$ [see Fig.~\ref{Fig:ZkDMRG}(b)]. Notably, $Z_\mathbf{k}$ at $t/J=-5$ is highly suppressed, thus we multiply it with a factor $30$ to clearly compare the two sides.
The finite $Z_\mathbf{k}$ and the ground-state momenta obtained from DMRG confirm the SCBA results, demonstrating the magnetic polaron theory indeed captures the nature of a doped charge.
The distinct momenta for opposite signs of $t/J$ also manifest the particle-hole asymmetry. Given the relation
$Z_\mathbf{k}\sim m/m^*$, with $m$ the bare electron mass and $m^*$ the effective mass of quasiparticle, our results indicate that the effective mass of the doped single charge at $t/J<0$ is much larger than $t/J>0$, or equivalently, the band at $t/J<0$ is quite narrow or nearly flat.

\begin{figure}[t]
    \centering
    \includegraphics[width=0.45\textwidth]{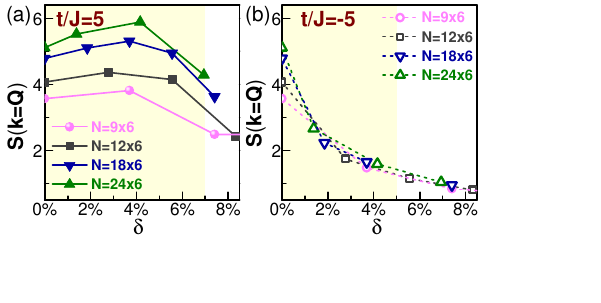}
    \renewcommand{\figurename}{Fig.}
    \caption{(Color online).
        The spin structure factor $S(\mathbf{k=Q})$ as a function of charge doping concentration $\delta$ for $t/J=5$ (a) and $t/J=-5$ (b) on $L_y=6$ cylinders. Results are from DMRG calculations.
    }
    \label{Fig:SqDMRG}
\end{figure}

To explore the distinct behavior of the doped charges at two sides and check the validity of our magnetic polaron theory, we probe the spin channel
by computing the static spin structure factor
\begin{equation}
S(\mathbf{k}) =\frac{1}{N}\sum_{i,j} {\left\langle {\mathbf{S}_i\cdot \mathbf {S}_j} \right\rangle e^{i \mathbf{k}\cdot (\mathbf{r}_i-\mathbf{r}_j)}}~,
\end{equation}
 which also can be directly probed  in an optical lattice \cite{Wurz2018}. At half filling, the $120^{\circ}$ N\'eel order is characterized by the sharp peaks of $S(\mathbf{k})$ at $\mathbf k=\mathbf{Q}$ [see Fig.~\ref{Fig:fig1}(c)].
To probe the evolution of the spin background with doping, we keep track of $S(\mathbf{k=Q})$ as a function of doping concentration $\delta$, as shown in Figs.~\ref{Fig:SqDMRG}.
For $t/J>0$, we find the spin background is insensitive to the light doping, as indicated by the robust sharp peak of $S(\mathbf{k=Q})$ for $\delta \lesssim 10\%$ [see Fig.~\ref{Fig:SqDMRG} (a)]. However, the $120^{\circ}$ N\'eel order is rapidly weakened at $t/J<0$ even with a much lower doping level $\delta\lesssim5\%$ [see Fig.~\ref{Fig:SqDMRG} (b)], suggesting the motion of the charge may induce a global distortion on the spin background, which in turn would further dress the doped charge, leading to the significantly enhanced effective mass. The $120^{\circ}$ N\'eel order is also the precondition of magnetic polaron theory, and our findings of $S(\mathbf{k})$ against doping suggest such a theory is valid at least for $\delta\lesssim5\%$ at both sides, while for a wider range at the $t/J>0$ side.
The distinct nature of the ground state in both charge and spin channels exhibits the particle-hole asymmetry and reveals the intricate interplay between the charge and spin degrees of freedom.

\section{Summary and Outlook}
 In summary, we theoretically predict asymmetric photoemission and tunneling spectroscopies for the lightly doped triangular Mott antiferromagnets and identify two distinct types of magnetic polarons on frustrated lattices: the lightly renormalized quasiparticle with the same momentum as the spin background, and the heavily renormalized quasiparticle with a shifted momentum, which resolves discrepancies in previous works \cite{PhysRevB53402,Trumper2004Quasiparticle,Vojta1999spinpolaron,PhysRevB72224409}.
The latter provides a new way to engineer the flat bands and explore the possible topology and Kondo physics in doped Mott insulators.
 We further show the asymmetric responses of the spin background against doping and confirm the validity of our theory at least within $\delta \lesssim5\%$.
Our findings might motivate future theoretical studies on the interplay between the local degrees of freedom and the geometric frustration, or on the possible emerged phases with further increased doping concentration \cite{Zhu2020}, both of which are of fundamental importance for understanding the Fermi-Hubbard physics on frustrated lattices. 

Moreover, the triangular-lattice Fermi-Hubbard model has recently been realized on the frustrated optical lattices~\cite{YangJin2021}, in which both the ratio $U/t$ and charge doping can be accurately tuned~\cite{Lewenstein2012,ChinCheng2010,Gross2021,Bohrdt2021, Bakr2009quantumgas, Parsons2016siteresolved, Sherson2010singleatomresolved, Cheuk2016Observation,YangJin2021,Schafer2020,Boll2016,Brown2017, Brown2019,GuardadoSanchez2020,Koepsell2020,Hartke2020}, then our predictions of the spectral function and the static spin structure factor are readily verified based on recently developed techniques including ARPES~\cite{2020NatPh1626B,Bohrdt2018} and the coherent manipulation of spin correlations~\cite{Wurz2018}.
Additionally, our prediction of the spectral function and the density of states
at light doping can also be directly tested in the TMD and its heterostructures by the STM and nano-ARPES techniques ~\cite{Tang2020Simulation,Kennes2021platform,Lisi2021}. These proposals based on our theory may also inspire more experimental ideas beyond the extensively studied square-lattice case.

\begin{acknowledgments}
We would like to thank Z. Y. Weng, F. C. Zhang, T. K. Ng, A. Vishwanath, D. N. Sheng, J. X. Li, T. Li,  J. H. Mao, Y. Xu, M. Knap for helpful discussions.
This work was supported by the National Natural Science Foundation of China (Grant No. 12074375), the Innovation Program for Quantum Science and Technology
(Grant No. 2-6), the Fundamental Research Funds for the Central Universities, the Strategic Priority Research Program of CAS (Grant No.XDB33000000) and the start-up funding of KITS at UCAS.
\end{acknowledgments}

\appendix

\section{Particle-hole transformations}

For the $t$-$J$ model [see Eq.~(2)] in the main text, the Hilbert space subject to the single-occupation condition and the single charge doping is realized by removing a fermion from the half-filled background, i.e., $\sum_{i}n_{i}=N-1$. We refer to the electron-doped and hole-doped physics as the cases of $t/J>0$ and $t/J<0$,  due to the particle-hole asymmetry. Here we give a more detailed explanation for it.

\begin{figure}[t]
\centering
\includegraphics[scale =0.4]{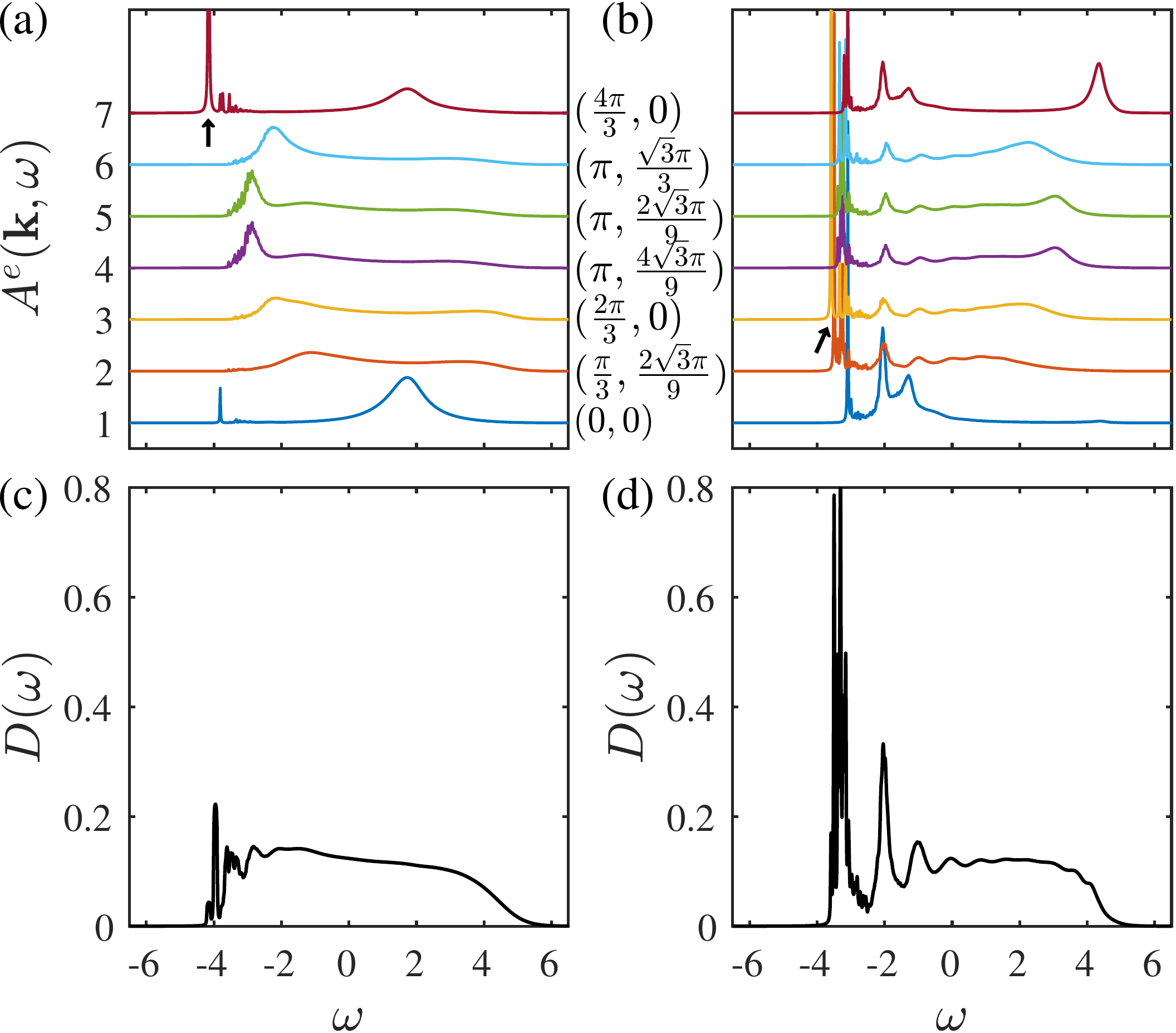}
\caption{(Color online) Spectra and density of states from the SCBA at $12\times12$~lattice size. A well-defined quasiparticle peak (marked by black arrows) lies at the bottom of spectrum with $\mathbf{K}=(\frac{4\pi}{3},0)$ for $t/J=5$ in (a) and $(\frac{2\pi}{3},0)$ for $t/J=-5$ in (b), respectively. The density of states $D(\omega)$ features a vanishing gap in a continuum limit for $t/J=5$ in (c) and  $t/J=-5$ in (d). Here we set $\vert t\vert=1$ as the unity of energy
and energy resolution $\Delta \omega=0.01.$}
 \label{Fig:Aw12}
\end{figure}

At half-filling $\sum_{i}n_{i}=N$, the Hubbard model in the large $U$ limit can be reduced to a pure Heisenberg model, which harbors the $120^{\circ}$ N\'eel order. For a single hole doping, i.e., $\sum_{i}n_{i}=N-1$, the low energy for the Hubbard model in Eq.~(1) in the main text can be well captured by the $t$-$J$ model with Hilbert space subject to the single-occupation condition. Similarly, for a single electron doping, i.e., $\sum_{i}n_{i}=N+1$, the perturbation mechanism will give the following low energy Hamiltonian,
\begin{equation}
H=-t\sum_{\left\langle ij\right\rangle \sigma}\widetilde{\mathcal{P}}(c_{i\sigma}^{\dagger}c_{j\sigma} +c_{j\sigma}^{\dagger}c_{i\sigma})\widetilde{\mathcal{P}}+J\sum_{\langle ij\rangle}\mathbf{S}_{i}\cdot\mathbf{S}_{j}~,
\label{eq:H1}
\end{equation}
where instead, the projector $\widetilde{\mathcal{P}}$ removes empty
occupancy. Subsequently, we introduce the particle-hole transformation
\begin{equation}
c_{i\sigma}\rightarrow c_{i\sigma}^{\dagger},c_{i\sigma}^{\dagger}\rightarrow c_{i\sigma}~,
\end{equation}
which changes the sign of the hopping integral in Eq.~(\ref{eq:H1}), i.e., $t\rightarrow-t$, but keeps $J$ invariant. Significantly, the particle number is mapped to
\begin{equation}
\sum_i n_i=N+1 \rightarrow  \sum_i n_i=N-1~.
\end{equation}
In this sense, different signs of the hopping integral $t$ refer to the electron-doped or hole-doped case for the $t$-$J$ model in Eq.~(2) in the main text.

\section{Finite Size effect}

\begin{figure}[b]
\centering
\includegraphics[scale =0.4]{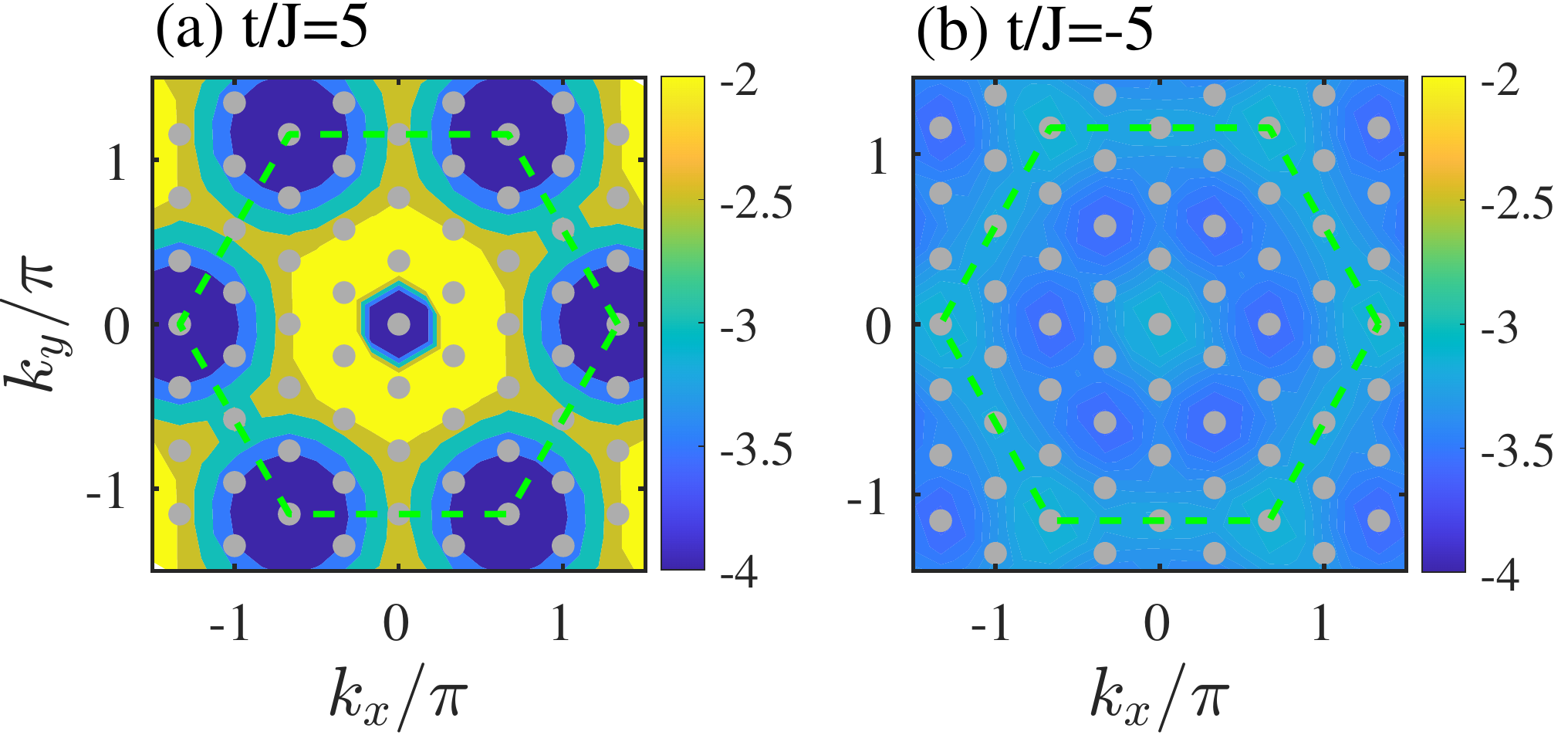}
\caption{(Color online) Dispersion for (a) $t/J=5$ and (b) $t/J=-5$. We can clearly see that the bandwidth for $t/J=5$ is much larger than that of $t/J=-5$. Here we set $\vert t\vert=1$ as the unity of energy and energy resolution $\Delta \omega=0.01$.}
\label{Fig: dispersion12}
\end{figure}

Here we present the results from the self-consistent Born approximation (SCBA) with a  larger lattice size.
Figure~\ref{Fig:Aw12} depicts the results calculated by SCBA with $12\times12$ lattice size, where the spectrum and densities of states are consistent with the $6\times6$ lattice. Furthermore, some noise like subtle structure gets suppressed. In Fig.~\ref{Fig:Aw12}(d), the extremely sharp peak indicates the Van Hove singularity behavior. In Fig.~\ref{Fig: dispersion12},
we plot the dispersion relations for $t/J=5$ in the left panel and
$t/J=-5$ in the right panel. Figure \ref{Fig: dispersion12}(b) for $t/J=-5$ shows a much narrower bandwidth, implied by the almost invariant color distribution with the same color scale as that of the $t/J=5$ side in Fig.~\ref{Fig: dispersion12}(a).

\section{The momentum distribution at light doping}

\begin{figure*}[th]
\centering
\includegraphics[width=1\textwidth]{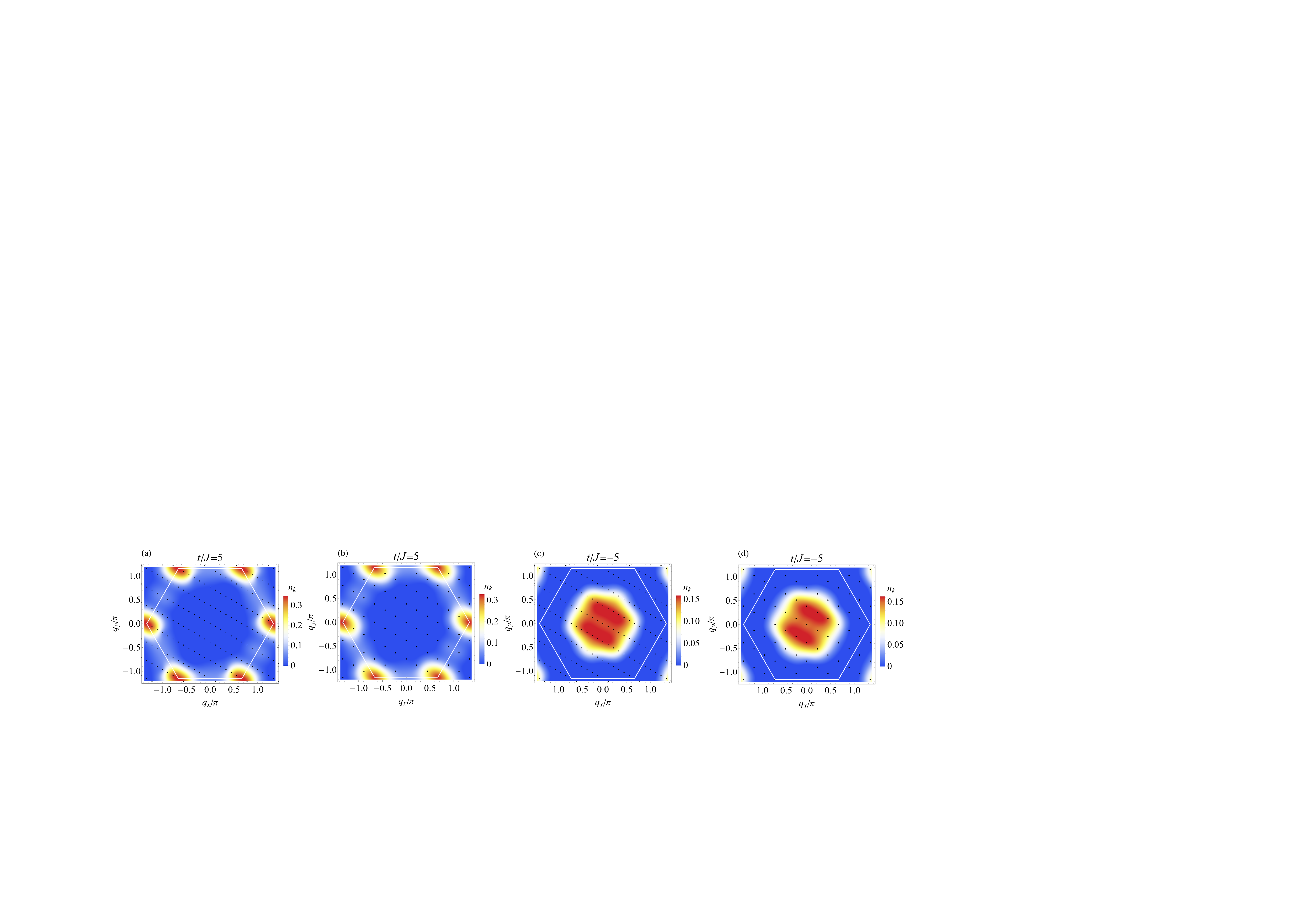}
\caption{(Color online) The momentum distribution of the doped charge $n^c_\mathbf{k}$ with doping concentration $\delta=1/27$. $t/J=5$ for (a,b) and $t/J=-5$ for (c,d), while $N=18\times 6$ for (a,c) and $N=9\times 6$ for (b,d).
The black dots represent the accessible momenta in the Brillouin zone (white lines). Interpolation has been applied.}
\label{fig:nk}
\end{figure*}

In this section, we examine the momentum distribution of the doped charge at  light doping.
The momentum distribution of the doped charge can be calculated by
\begin{equation}
 n^c_{\mathbf{k}} \equiv 1-\sum_{\sigma}\left\langle\Psi_{G}\left|c_{\mathbf{k} \sigma}^{\dagger} c_{\mathbf{k} \sigma}\right| \Psi_{G}\right\rangle,
\end{equation}
where $|\psi_G\rangle$ is the ground state, and $c_{\mathbf{k} \sigma}^{\dagger}$ ($c_{\mathbf{k} \sigma}$) is the fermion creation (annihilation) operators with momentum $\mathbf{k}$ and spin $\sigma=\uparrow,\,\downarrow$. The factor 1 corresponds to the momentum distribution at half filling for a Mott insulator.

\begin{figure}[b]
\centering
\includegraphics[width=0.5\textwidth]{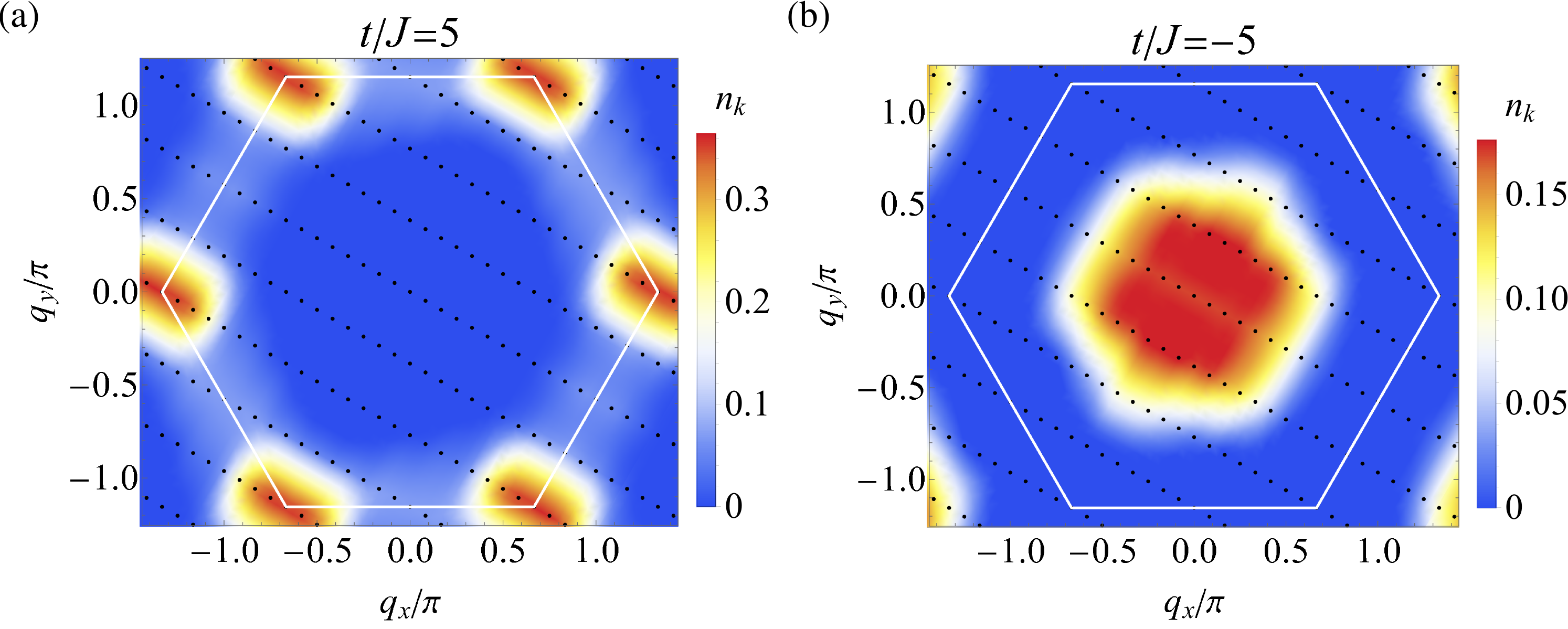}
\caption{(Color online) The momentum distribution of the doped charge $n^c_\mathbf{k}$ with doping concentration $\delta=1/24$. $t/J=5$ for (a) and $t/J=-5$ for (b) on $N=24\times 6$ lattice. The black dots represent the accessible momenta in the Brillouin zone (white lines). Interpolation has been applied.}
\label{fig:nk24}
\end{figure}

In Fig.~\ref{fig:nk}, we show the momentum distribution of the doped charge $n^c_\mathbf{k}$  at finite doping on lattices with two different sizes $N=9\times6$ and $N=18\times6$. We find the physical indications are robust against system size.
As shown in Figure~\ref{fig:nk}(a) or (b) for $t/J=5$, we find that, with the increase of the doping concentration, the Fermi pockets are gradually formed around $\mathbf{K}$ points compared with the single charge momentum $\mathbf{K}$ [see Fig.~2(a) and Fig.~3(a) in the main text]. 
In particular,  since the momentum is a discrete value $(2\pi/L_i)\cdot m_i$ in the direction $i$, where $m_i\in\mathbb{Z}$ and $i=x,\,y$ with the finite system size $L_x$ and $L_y$, the value of the Fermi momentum is a size-dependent quantity. However, the numerical observations of the formation of Fermi pockets around $\mathbf{K}$ are consistent with such quasiparticle behavior. By contrast,  for the other side $t/J=-5$, as shown in Fig.~\ref{fig:nk}(c,d), the doped charge forms a large Fermi surface with a wider distribution of the Fermi momentum, as indicated by the sudden drop of $ n^c_{\mathbf{k}}$. The large Fermi surface suggests that many low-energy quasiparticles with different momenta share close energies, which is consistent with filling the nearly flat band that emerges for the single-charge doping case at $t/J<0$, as shown in Fig.~2(b) and Fig.~3(b) in the main text.

We further confirm the above interpretations by slightly increase the doping concentrations from $\delta=1/27$ [see Fig.~\ref{fig:nk}] to $\delta=1/24$ [see Fig.~\ref{fig:nk24}], and find the numerical observations are consistent with the physical picture mentioned above.

\section{Details for the SCBA calculations}
At half-filling, the strong Hubbard interaction penalizes double occupancy
at sites and thus the ground state harbors well-established $120^{\mathrm{\circ}}$
antiferromagnetic collinear order that is described by a pure Heisenberg
model $H_{J}=J\sum_{\left\langle ij\right\rangle }\mathbf{S}_{i}\cdot\mathbf{S}_{j}$.

We can apply the spin wave theory to characterize the low energy excitations.
Without loss of generality, we set the AFM order residing the XZ-plane
by rotating each spin by an angle $\mathbf{Q}\cdot\mathbf{r}_{i}$
along y-axis at site $i$
\begin{align}
H_{J}\rightarrow & -\frac{J}{2}\sum_{\left\langle ij\right\rangle }S_{i}^{z}S_{j}^{z}+J\sum_{\left\langle ij\right\rangle }\left(\frac{1}{2}S_{i}^{x}S_{j}^{x}+S_{i}^{y}S_{i}^{y}\right)  \notag \\
&+\sin\mathbf{Q}\cdot\mathbf{r}_{ij}\left(S_{i}^{x}S_{j}^{z}-S_{i}^{z}S_{j}^{x}\right)~,
\label{eq:Hrot}
\end{align}
where $\mathbf{r}_{ij}\equiv\mathbf{r}_{i}-\mathbf{r}_{j}$ and $\mathbf{Q}$
is the momentum of spin order. The second term in Eq.~(\ref{eq:Hrot})
is from geometric frustrations, with no counterpart in a square lattice.
The magnon polarons can be described by the HP bosons (in the large
$S$ limit)
\begin{equation}
S_{i}^{z}=S-a_{i}^{\dagger}a_{i},S_{i}^{+}=a_{i},S_{i}^{-}=a_{i}^{\dagger}~,
\label{eq:HPbosons}
\end{equation}
Here $S_{i}^{\pm}=S^{x}\pm iS^{y}$ is the spin ladder operators.
To the leading order, Eq.~(\ref{eq:HPbosons}) yields an effective
Hamiltonian for magnons,
\begin{equation}
H_{a}=\frac{J}{2}\sum_{i}2S\nu n_{i}+\frac{SJ}{4}\left(-3a_{i}a_{j}+a_{i}a_{j}^{\dagger}+a_{i}^{\dagger}a_{j}-3a_{i}^{\dagger}a_{j}^{\dagger}\right)~.
\label{eq:Ha}
\end{equation}
The last term in Eq.~(\ref{eq:Hrot}) involves higher-order magnon
interactions and can be safely neglected. We can diagonalize $H_{a}=\sum_{k}\frac{\nu}{2}JS\omega_{\text{\ensuremath{\mathbf{k}}}}^{s}\mathbf{\beta_{k}^{\dagger}}\beta_{\mathbf{k}}$
in Eq.~(\ref{eq:Ha}) by the Bogoliubov transformation
\begin{align}
\beta_{\mathbf{k}} & =u_{\mathbf{k}}a_{\mathbf{k}}-\upsilon_{\mathbf{k}}a_{-\mathbf{k}}^{\dagger}~,\\
\beta_{-\mathbf{k}}^{\dagger} & =-\upsilon_{\mathbf{k}}a_{\mathbf{k}}+u_{\mathbf{k}}a_{-\mathbf{k}}^{\dagger}~,
\end{align}
where $\gamma_{\mathbf{k}}=\frac{1}{\nu}\sum_{\delta}e^{i\mathbf{k}\cdot\delta}$
sums over all $\nu=6$ nearest-neighbor sites. Here $\omega_{\mathbf{k}}^{a}=\sqrt{\left(1-\gamma_{\mathbf{k}}\right)\left(1+2\gamma_{\mathbf{k}}\right)}$
and $\omega_{\mathbf{k}}^{s}=\frac{\nu}{2}JS\omega_{\mathbf{k}}^{a}$
is the magnon's dispersion. The coefficients $u_{\mathbf{k}}$and $\upsilon_{\mathbf{k}}$
are the usual Bogoliubov factors in the linear spin-wave theory
\begin{equation}
u_{\mathbf{k}}=\sqrt{\frac{1+\gamma_{\mathbf{k}}/2+\omega_{\mathbf{k}}^{a}}{2\omega_{\mathbf{k}}^{a}}},\quad \upsilon_{k}=\mathrm{\mathop{sign}(\gamma_{\mathbf{k}})\sqrt{\frac{1+\gamma_{\mathbf{k}}/2-\omega_{\mathbf{k}}^{a}}{2\omega_{\mathbf{k}}^{a}}}}~.
\end{equation}
The half-filled ground state after rotation accordingly can assume
the form as
\begin{equation}
\vert\Psi_{0}\rangle=\exp\left(-\sum_{\mathbf{k}}\frac{v_{\mathbf{k}}}{u_{\mathbf{k}}}a_{\mathbf{k}}a_{-\mathbf{k}}\right)\vert\text{N\'eel}\rangle~,
\label{eq:quantumNeel}
\end{equation}
with $|\text{N\'eel}\rangle$ being a classical $120^{\circ}$ AFM
order.
The SCBA method considers the propagation of a holon excitation by
absorbing or emitting magnons $a_{i}$ on a quantum N\'eel ground state
in Eq.~(\ref{eq:quantumNeel}). On the rotated state $\vert\Psi_{0}\rangle$,
we assume a spinless holon $h_{i}^{\dagger}$ is to be created by
removing a spin-$\uparrow$ electron. Namely, we have a typical fractionalization
scheme,
\begin{equation}
c_{i\uparrow}=h_{i}^{\dagger},c_{i\downarrow}=h_{i}^{\dagger}S_{i}^{+}=h_{i}^{\dagger}a_{i}~,\label{eq:fractionalization}
\end{equation}
By ignoring higher-order interactions involving multi-magnons, the
hopping terms will split into two terms: the kinetic energy term for
the holon,
\begin{equation}
H_{h0}=-\sum_{\mathbf{k}}\omega_{0}^{h}(\mathbf{k})h_{\mathbf{k}}h_{\mathbf{k}}^{\dagger}~,
\label{eq:Hh0}
\end{equation}
 with a bare dispersion $\omega_{0}^{h}(\mathbf{k})=\frac{\nu t}{2}(\gamma_{\mathbf{k}+\frac{Q}{2}}+\gamma_{\mathbf{k}-\frac{Q}{2}})$
and the holon-magnon coupling terms
\begin{equation}
H_{hb}=-\frac{\nu t}{i\sqrt{N}}\sum_{\mathbf{k}}h_{\mathbf{k}}h_{\mathbf{k}-\mathbf{q}}^{\dagger}(M_{\mathbf{k}}a_{\mathbf{q}}^{\dagger}-M_{\mathbf{k-q}}a_{-\mathbf{q}})~,
\end{equation}
with $M_{\mathbf{k}}=\gamma_{\mathbf{k}+\mathbf{Q}/2}-\gamma_{\mathbf{k}-\mathbf{Q}/2}$.
The existence of the kinetic term $H_{h0}$ in Eq.~(\ref{eq:Hh0}),
which is absent in the square lattice, indicates the Trugman loop may not play a leading role, in contrast to the square lattice. The holon-magnon interaction
$H_{hb}$ opens a channel for propagation of a doped charge by absorbing
or emitting magnons. The holon's Green's function with a definition $G^{h}(\mathbf{k},\omega)=\langle\Psi_{0}\vert h_{k}\frac{1}{\omega-H}h_{k}^{\dagger}\vert\Psi_{0}\rangle$,
will be determined a self-consistent equation by only involving
non-crossed Feynman diagrams for the holon's self energy $\Sigma^{h}(\mathbf{k},\omega)$,
\begin{equation}
\Sigma^{h}(\mathbf{k},\omega)=\sum_{\mathbf{q}}\frac{f(\mathbf{k},\mathbf{q})}{\omega-\omega_{0}^{h}(\mathbf{k}-\mathbf{q})-\omega_{\mathbf{q}}^{s}-\Sigma^{h}\left(\mathbf{k}-\mathbf{q},\omega-\omega_{\mathbf{q}}^{s}\right)}~,
\label{eq:self_energy}
\end{equation}
with a vertex coupling,
\begin{equation}
f(\mathbf{k},\mathbf{\mathbf{q}})=\frac{1}{N}(\nu t)^{2}\vert M_{\mathbf{k}}\upsilon_{\mathbf{q}}-M_{\mathbf{k}-\mathbf{q}}u_{\mathbf{q}}\vert^{2}~.
\end{equation}
The experimentally detectable electron's Green's function $G^{e}(\mathbf{k},\omega)$
is related based the fractionalization scheme in Eq.~(\ref{eq:fractionalization}),
\begin{equation}
G^{e}(\mathbf{k},\omega)=-\frac{1}{4}G^{h}(\mathbf{k}+\frac{\mathbf{Q}}{2},\omega)-\frac{1}{4}G^{h}(\mathbf{k}-\frac{\mathbf{Q}}{2},\omega)~,
\end{equation}
where $\mathbf Q/2$ arises from the rotating operations on the spins at each site.
The self-consistent equation for the self-energy $\Sigma^{h}$ can
be tackled by numerical iterations on randomly initialized values of
$\Sigma^{h}(\mathbf{k},\omega)$. At the side of $t<0$, one needs
a large number of iterations to reach the convergence, which may be
traced back to the existence of the nearly flat band and the extremely weak quasiparticle
weight.

\begin{figure}[t]
    \centering
    \includegraphics[scale=0.33]{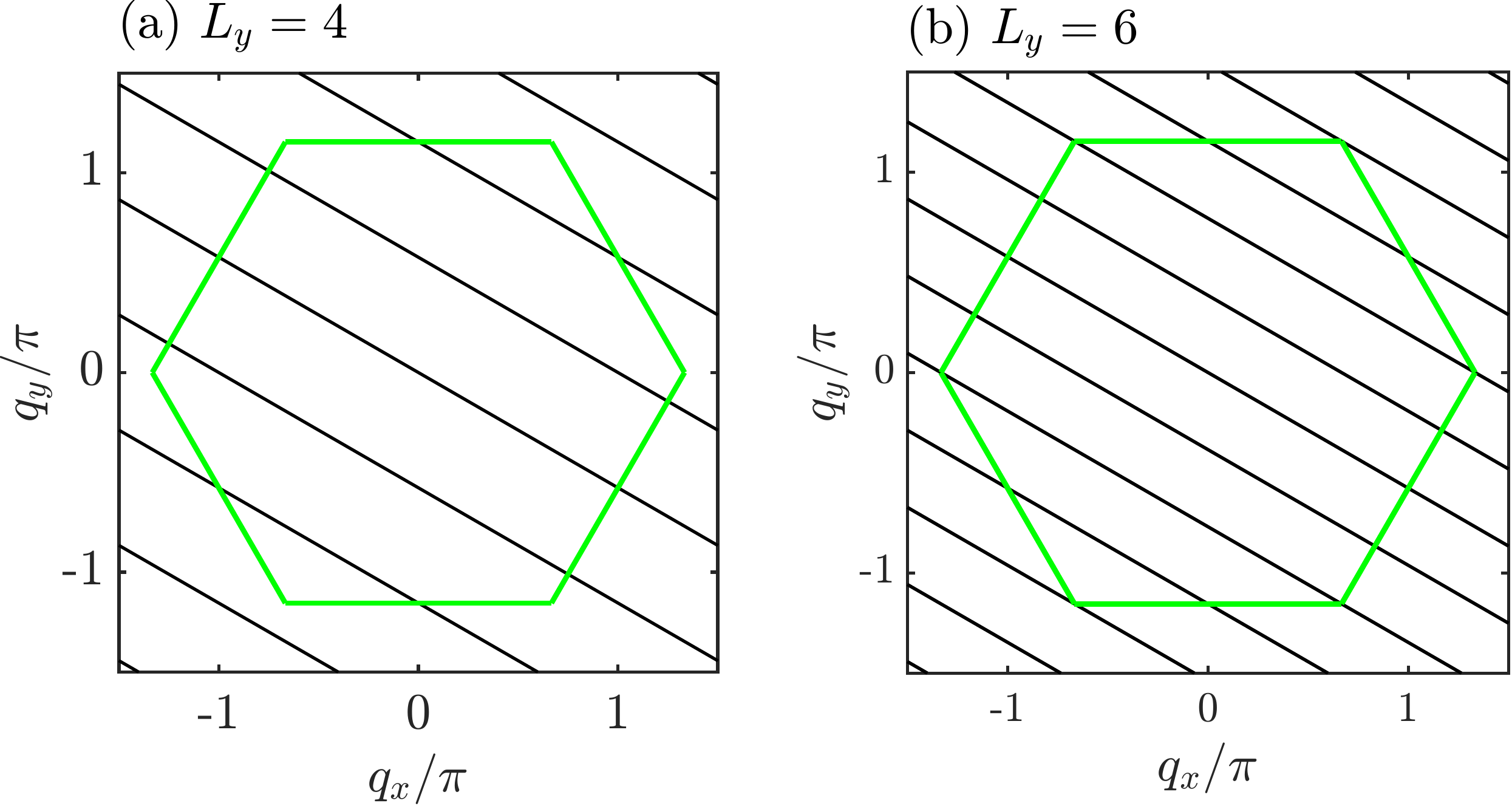}
    \caption{
    Finite size effects for accessible momenta marked by black lines in the Brillouin zone: (a) four-leg and (b) six-leg ladders. The boundary of the first Brillouin zone is guided by a green hexagon.
    }
    \label{Fig:finitesize}
\end{figure}

\section{Comparison with previous works}

The fate of the injected charges in the Fermi-Hubbard model on a triangular lattice is a long-standing issue but with controversial conclusions. In previous literature, no consistent results are reached, even for those using the same SCBA method  and exact diagonalization method.
Here we give a detailed illustration.

Ref.~\cite{PhysRevB53402} studied the single-charge problem on a triangular lattice. Its predictions on the ground state momenta are peaked at $\mathbf K$ and $\mathbf K/2$ respectively for $t/J=\pm 5$, which is consistent with ED on a comparatively small lattice.
However, controversies were erupted against this prediction with ED on a larger lattice in Ref.~\cite{PhysRevB69184407}.
Ref.~\cite{PhysRevB69184407} revisited the problem with the same SCBA method, but affirms the absence of quasiparticle excitations outside the neighborhood of magnetic Goldstone modes at the side $t/J=5$, the ground state momentum resides $\mathbf K$ for $t/J=5$ and $\mathbf \Gamma$ for $t/J<-5$~\cite{PhysRevB69184407},
which is consistent with ED on a larger lattice.
Concomitantly, other methods report discrepancies. For example,
Ref.~\cite{PhysRevB596027} adopted a cumulant version of Mori-Zwanzig projection techniques
to deal with spin bag quasiparticle or magnetic polaron,
and found the ground-state momenta locates
in $\mathbf K$ points at both sides $t>0$ and $t<0$.
The discrepancies motivate our works, in particular by applying the
advanced DMRG method to avoid the severe finite size effect in ED.
For example, at $6\times 4$ lattice size, both K and K/2 are not accessible (see \ref{Fig:finitesize}(a)) and on the width-three ladder, the spin background has a finite gap due to the spontaneous dimerization. The minimum lattice size to capture the $120^\circ$ order in the absence of dimerization is $6\times6$ (see \ref{Fig:finitesize}(b)), which is beyond the capacity of ED.

%

\end{document}